\documentclass[amsmath, aps]{revtex4}
\usepackage{graphicx}
\usepackage{times}
\usepackage{epsfig}
\usepackage{amssymb}
\usepackage{amsmath}

\newcommand{\hr}{\hat{\bf r}}
\begin{document}
\title{Two-photon decays of highly excited states in hydrogen}

\author{D. Solovyev$^1$, V. Dubrovich$^2$, A. V. Volotka$^{1,3}$,
L. Labzowsky$^{1,4}$, and G. Plunien$^3$}

\affiliation{ 
$^1$ V. A. Fock Institute of Physics, St. Petersburg
State University, Petrodvorets, Oulianovskaya 1, 198504,
St. Petersburg, Russia
\\
$^2$ St. Petersburg Branch of Special Astrophysical Observatory,
Russian Academy of Sciences, 196140, St. Petersburg, Russia
\\
$^3$ Institut f\"ur Theoretische Physik, Technische Universit\"at Dresden,
Mommsenstrasse 13, D-01062, Dresden, Germany
\\
$^4$  Petersburg Nuclear Physics Institute, 188300, Gatchina, St. Petersburg, Russia
}

\begin{abstract}
The relativistic and nonrelativistic approaches for the calculations
of the two-photon decay rates of highly excited states in hydrogen
are compared. The dependence on the principal quantum number ($n$) of
the $ns$, $nd$, and $np$ initial states is investigated up to $n=100$
for the nonresonant emissions. For the $ns$ states together with the main
E1E1 channel the contributions of higher multipoles (M1M1, E2E2, E1M2)
are considered. For the $np$ states the E1M1 and E1E2 channels are evaluated.
Moreover, the simple analytical formula for the E1M1 decay is derived in the
nonrelativistic limit.
\end{abstract}
\maketitle

\section{Introduction}
The question about the role of the two-photon decays in the context of the hydrogen
recombination in early Universe was raised at first in \cite{Zeldovich,Peebles}. 
It was shown that the electron recombination occurs mainly into the $2s$ state,
followed by the $2s$ state decays via two-photon E1E1 emission into the ground state.
The transition frequencies of these photons are not resonant
(the energy of each photon released will not excite any of neighbouring atom),
and, therefore, the medium is transparent for these frequencies.
The contribution of such transition rate to the microwave background was further
investigated in the literature, see, e.g., \cite{seager2000,dubrovich2005,chluba2006}.
The recent success in observation of the cosmic microwave background temperature
and polarization anisotropy draws attention to the details of cosmological
hydrogen recombination history. This, in turn, requires an accurate knowledge of
the two-photon decay processes.

The theoretical formalism for the two-photon decay has been introduced by
G\"oppert-Mayer \cite{Goeppert} and the first calculation of the two-photon
E1E1 transition $2s \rightarrow 2\gamma(E1)+1s$ belongs to Breit and Teller
\cite{Breit}. Further improvements including relativistic corrections
were performed in works \cite{johnson1972,goldman1981}. The two-photon
transition $2p - 1s$ occurring via E1E2 and E1M1 channels was
first evaluated in \cite{labzowsky2005,labzowsky2006}.
In recent paper by Amaro {\it et al.} \cite{amaro2009} these results were
confirmed and further two-photon transition rates from initial bound states
with $n = 2,3$ have been evaluated.

In connection with increased accuracy of the astrophysical experiments
(see, e.g., \cite{wong2008}) it might become also important to consider atomic
recombination into highly excited states in hydrogen, as proposed by Dubrovich
and Grachev in \cite{dubrovich2005}, and further developed in
\cite{chluba2008,hirata2008,chluba2010}.
The magnitude of the nonresonant radiation from highly excited states
is of the same order as for the $2s-1s$ transition and, therefore,
the same contribution to the emission escape can be anticipated.
Highly excited states can decay by cascade emission (i.e, the transition occurs via
intermediate state) and by ``pure'' (nonresonant) two-photon
emission. In the case of the cascade transition the emitted photons can be absorbed
immediately by the neighbouring atoms and, therefore, the medium is not
transparent for the resonant radiation, but it is for the nonresonant one.
For this reason the question of separation of the ``pure'' and cascade emission
arises. There are some papers considering this problem (see, e.g.,
\cite{jentschura2009,wundt2009,labzowsky2009,jentschura2010}).
In the papers \cite{jentschura2009,wundt2009} the possibility of
separation between cascade and ``pure'' two-photon emission has been shown.
However, in our paper \cite{labzowsky2009} it has been demonstrated that it is
impossible to divide resonant and nonresonant two-photon emission due to the
interference term between them. Recently, in \cite{jentschura2010} Jentschura
accepted our point of view on the treatment of this problem.
The separation of the cascades and the ``pure'' two-photon emission remains
ambiguous. In works \cite{chluba2008,chluba2010} the influence of the highly excited
states to the Sobolev escape probability for Lyman-$\alpha$ photons was computed,
employing the ``1+1'' scheme. In the vicinity of the Lyman-$\alpha$ resonance
the shape of the ``1+1'' emission profile remains Lorentzian and the
resonant/nonresonant or resonant/resonant interference terms do not appear.
Another approach for the account of the two-photon processes from the highly excited
states to the recombination problem was presented by Hirata in \cite{hirata2008}
introducing a distinction between regions with resonant photon contributions and
those with ``pure'' two-photon contributions.

The main aim of the present paper is to evaluate rigorously the dependence of the
two-photon decay rates on the principal quantum number of initial
state $n$. Since the nonresonant photons being transparent for the medium
contributes to the microwave background, it is important to investigate the
$n$-behaviour of the ``pure'' two-photon emission.
For this reason we calculate the differential transition rate for different
$n$ at the equal frequencies of the emitted photons, the energy region where
the contribution of the cascades is almost negligible for high $n$.
In the calculations we employ both the relativistic and nonrelativistic approaches.
By the nonrelativistic approach we mean: (i) the long-wavelength approximation
for the photon emission operator, in the case of dipole transitions
this yields the dipole operators, therefore we will use the term
``dipole approximation'' for this; (ii) matrix elements evaluated with
the nonrelativistic (Schr\"odinger) electron wave functions.
In the relativistic approach the full photon emission operator together with
the relativistic electron wave functions are employed.
Since the orbital size of the highly excited states increases quadratically
with the increasing of $n$, the long-wavelength (dipole) approximation
seems to be inadequate \cite{dubrovich2005}.
The argument of the photon wave function is no longer small and, therefore,
the restriction to the first term only in the power-series expansion of
the Bessel function is under the question. One could expect that the utilization
of the full photon emission operator contributes essential to the two-photon decay rates
from the highly excited states. For the same reason the higher multipoles
contributions (besides the dominant E1E1 channel) of the two-photon transitions
for the $ns$ and $nd$ levels could become important.
Moreover, since the highly excited states are strongly mixed, we should
consider, the decays from the $np$ states as well. 

The paper is organized as follows: In Sec.~II we present the general formulas
for the two-photon decay. In Sec.~III we calculate E1E1 differential transition
rates $ns/nd \rightarrow 1s$ at the equal photon frequencies in hydrogen
atom. The influence of the higher multipoles (M1M1, E2E2, and E1M2) is also
investigated. As next step we consider the two-photon E1E2 and E1M1 transition
rates of the $np$ states (Sec.~IV and V). The investigation of the dependence on
$n$ is performed for all these transitions.
Moreover, in Sec.~V we derive simple analytical formula for the E1M1 decay rates of
the $np$ states in the nonrelativistic limit. 

Apart from Sec.~II and III where the relativistic units are useful
for convenience, the atomic units are used throughout the paper.

\section{General formulas: two-photon decay}
The two-photon transition probability $A\rightarrow A'+2\gamma$ corresponds
to the following second-order $S$-matrix elements
\begin{eqnarray}
\label{18}
\langle A'|\hat{S}^{(2)}|A\rangle  = e^2\int d^4x_1d^4x_2
\left(\bar{\psi}_{A'}(x_1)\gamma_{\mu_1}A^*_{\mu_1}(x_1)S(x_1x_2)
      \gamma_{\mu_2}A^*_{\mu_2}(x_2)\psi_A(x_2)\right)\,,
\end{eqnarray}
where $S(x_1x_2)$ is the Feynman propagator for the atomic electron electron. 
In the Furry picture the eigenmode decomposition for this propagator reads
(e.g. \cite{Akhiezer}, see also \cite{labzowsky2009})
\begin{eqnarray}
\label{19}
S(x_1x_2)=\frac{1}{2\pi i}\int\limits_{-\infty}^{\infty}
d\omega\,e^{i\omega(t_1-t_2)}\sum\limits_N\frac{\psi_N(\vec{r}_1)
\bar{\psi}_N(\vec{r}_2)}{E_N(1-i0)+\omega}\,,
\end{eqnarray}
where the summation in Eq. (\ref{19}) extends over the entire
Dirac spectrum of electron states $N$ in the field of the nucleus,
$\psi_N(x)$ is the electron wave function,
$E_N$ is the electron energy. In Eq. (\ref{18}) $\gamma_{\mu}$
are the Dirac matrices. The wave function of the photon 
\begin{eqnarray}
\label{u10}
A^{\vec{k},\lambda}_{\mu}(x)=\sqrt{\frac{2\pi}{\omega}}
e^{(\lambda)}_{\mu}e^{i(\vec{k}\vec{r}-\omega t)}
\end{eqnarray}
is characterized by the momentum $\vec{k}$ ($\omega=|\vec{k}|$)
and polarization vector $e_{\mu}^{(\lambda)}$
($\mu,\lambda=0,1,2,3$), $x\equiv(\vec{r},t)$. For real
transverse photons we have
\begin{equation}
\label{u28}
\vec{A}(x)=\sqrt{\frac{2\pi}{\omega}}\vec{e}
e^{i(\vec{k}\vec{r}-\omega t)}\equiv
\sqrt{\frac{2\pi}{\omega}}\vec{A}_{\vec{e},\vec{k}}
(\vec{r})e^{-i\omega t}\,.
\end{equation}
Integrating over time and frequency variables, taking into
account photon permutation symmetry and introducing the amplitude $U_{A'A}$ as
\begin{equation}
\begin{array}{l}
S_{A'A}^{(2\gamma)}=-2\pi
i\delta(E_{A'}+\omega+\omega'-E_A)U_{A'A}^{(2\gamma)}\,,
\end{array}
\end{equation}
yields the expression
\begin{equation}
\label{u11}
U^{(2\gamma)}_{A'A}=\frac{2\pi e^2}{\sqrt{\omega\omega'}}\left\{
\sum_N\frac{(\vec{\alpha}\vec{A}^*_{\vec{e},
\vec{k}})_{A'N}(\vec{\alpha}\vec{A}^*_{\vec{e}\,',
\vec{k}'})_{NA}}{E_N-E_A+\omega'}
+\sum_N\frac{(\vec{\alpha}\vec{A}^*_{\vec{e}\,',\vec{k}'})_{A'N}
(\vec{\alpha}\vec{A}^*_{\vec{e},\vec{k}})_{NA}}{E_N-E_A+\omega}
\right\}\,,
\end{equation}
where $e$ is the electron charge, $\vec{\alpha}$ is the vector incorporating
the Dirac matrices. The labels $A$, $A'$ and $N$ abbreviate the complete set of
the atomic electron quantum numbers (principal quantum number $n$, total angular
momentum $j$, its projection $m$, and parity $l$) for the initial, final and
intermediate states, respectively.
The transition probability is defined via
\begin{eqnarray}
\label{u12}
dW^{(2\gamma)}_{AA'}=2\pi\delta(E_A-E_{A'}-\omega-\omega')\left|U^{(2\gamma)}_{A'A}
\right|^2\frac{d\vec{k}}{(2\pi)^3}\frac{d\vec{k}'}{(2\pi)^3}\,.
\end{eqnarray}
Setting $d\vec{k}\equiv\omega^2d\vec{\nu}d\omega$ and integrating over $\omega$ yields
\begin{eqnarray}
\label{u13}
dW_{AA'}^{(2\gamma)}(\omega',\vec{\nu},\vec{\nu}\,',
\vec{e},\vec{e}\,')=e^4
\frac{\omega'(E_A-E_{A'}-\omega')}{(2\pi)^3}\sum\limits_{m_{A}m_{A'}}\frac{1}{2j_{A}+1}
\nonumber\\
\times\left|\sum_N\frac{({\vec{\alpha}}\vec{A}^*_{\vec{e},
\vec{k}})_{A'N} ({\vec{\alpha}}\vec{A}^*_{\vec{e}\,',
\vec{k}'})_{NA}}{E_N-E_A+\omega'}+\sum_N\frac{({\vec{\alpha}}\vec{A}^*_{\vec{e}\,',
\vec{k}'})_{A'N} ({\vec{\alpha}}\vec{A}^*_{\vec{e},
\vec{k}})_{NA}}{E_N-E_A+\omega}\right|^2d{\vec{\nu}}
d\vec{\nu}\,'d\omega'\,.
\end{eqnarray}
Here the summation over $N$ represents the summation over all sets of
quantum numbers of the intermediate state $\{nljm\}$. The sums over
the projections of the total angular momentum  of the
final state $A'$ and the averaging over the projections of the total
angular momentum of the initial state $A$ in Eq. (\ref{u13}) are also
included.

Expanding further the photon wave function in a multipole series
(see, e.g., Refs. \cite{Akhiezer,johnson1995}) we consider contributions
with different multipole structure E1E1, E1M1, E1E2, E1M2, M1M1, and E2E2.
The nonrelativistic limit can be easily obtained from
Eqs. (\ref{u28}) and (\ref{u13}), replacing
the photon and electrons wave functions by their nonrelativistic analogue.
In this way we describe electric dipole, magnetic dipole and electric quadrupole
photons (see, e.g., Ref.~\cite{solovyev2010}).

\section{E1E1 decay of the $ns$, $nd$ states}
First, we focus on the decay rate of the $ns$ and $nd$ levels 
($A\equiv ns/nd \rightarrow A'\equiv 1s$) in hydrogen. 
The common expression for the two-photon transition probability in relativistic evaluation
can be written as
\begin{eqnarray}
dW_{AA'}^{(2\gamma)}(\omega) = e^4
\frac{32\pi\,\omega\,\omega'}{2j_{A}+1}\sum\limits_{J\,J'\,\lambda\,\lambda'}
\;\sum\limits_{M\,M'\,m_{A}\,m_{A'}}
\Biggl|\sum_N\frac{(A^{(\lambda)}_{JM})_{A'N}
(A^{(\lambda')}_{J'M'})_{NA}}{E_N-E_A+\omega'}
+\sum_N\frac{(A^{(\lambda')}_{J'M'})_{A'N}
(A^{(\lambda)}_{JM})_{NA}}{E_N-E_A+\omega}\Biggr|^2 d\omega\,,
\end{eqnarray}
where the summation over the photon polarization and the integration
over the photon angles have been carried out, $\omega' = E_A-E_{A'}-\omega$.
Here $A^{(\lambda)}_{JM}(\omega)$ are the multipole components of the
transition operator, the symbol $\lambda$ indicates magnetic multipoles
($\lambda=0$) or electric multipoles ($\lambda=1$). The multipole components
$A^{(\lambda)}_{JM}(\omega)$ are defined in the same way as in
\cite{Akhiezer,johnson1995,volotka2002}.
They are given in the transverse gauge by the expressions
\begin{eqnarray}
\label{phot_vel}
A^{(0)}_{JM}(\omega)&=&j_J(\omega r)\ \vec{\alpha}\vec{Y}^{(0)}_{JM}(\hr)
\, ,\nonumber \\
A^{(1)}_{JM}(\omega)&=&\left(j'_J(\omega r)+\frac{j_J(\omega r)}{\omega r}\right)
\vec{\alpha}\vec{Y}^{(1)}_{JM}(\hr)+\sqrt{J(J+1)}\frac{j_J(\omega r)}{\omega r}
\ \vec{\alpha}\vec{Y}^{(-1)}_{JM}(\hr)\,,
\end{eqnarray}
where $j_J(x)$ is the spherical Bessel function, $\vec{Y}^{(\lambda)}_{JM}(\hr)$
are the vector spherical harmonics. The transverse gauge is also called velocity gauge
since the electric dipole matrix elements in this gauge turn into the velocity form
dipole amplitudes in the nonrelativistic limit. In the length gauge the magnetic
multipole components have the same form, whereas the electric multipole components
are given by
\begin{eqnarray}
\label{phot_len}
A^{(1)}_{JM}(\omega)&=&
-j_{J+1}(\omega r)\ \vec{\alpha}\vec{Y}^{(1)}_{JM}(\hr)
+\sqrt\frac{J+1}{J}j_{J+1}(\omega r)\ \vec{\alpha}\vec{Y}^{(-1)}_{JM}(\hr)
\nonumber \\
&&-i\sqrt\frac{J+1}{J}j_J(\omega r)Y_{JM}(\hr){\rm I}\,,
\end{eqnarray}
where ${\rm I}$ is the identity operator, $Y_{JM}(\hr)$ are the spherical functions.
Explicit formulas for the one-electron matrix elements $A^{(\lambda)}_{JM}(\omega)$
in the length and velocity gauges can be found in \cite{johnson1995}.
For the case of E1E1 transition one can easily obtain
\begin{eqnarray}
\label{21}
dW_{AA'}^{\rm E1E1}(\omega) = e^4
\frac{32\pi\,\omega\,\omega'}{2j_{A}+1}
\sum\limits_{M\,M'\,m_{A}\,m_{A'}}
\Biggl|\sum_N\frac{(A^{(1)}_{1M})_{A'N}
(A^{(1)}_{1M'})_{NA}}{E_N-E_A+\omega'}
+\sum_N\frac{(A^{(1)}_{1M'})_{A'N}
(A^{(1)}_{1M})_{NA}}{E_N-E_A+\omega}\Biggr|^2 d\omega\,.
\end{eqnarray}

The nonrelativistic limit for the E1E1 transitions can be obtained by
expanding the Bessel functions in the photon wave functions
Eqs.~(\ref{phot_vel})-(\ref{phot_len}) and keeping only the
nonrelativistic value for the large component of the electron wave functions.
Thus, we come to the following expressions in the length gauge
\begin{eqnarray}
\label{22}
dW^{\rm E1E1}_{ns,1s}(\omega)=e^4\frac{8\,\omega^3\,\omega'^3}{27\pi}
\left|S_{1s,ns}(\omega)+S_{1s,ns}(\omega')\right|^2d\omega
\end{eqnarray}
and
\begin{eqnarray}
\label{23}
dW^{\rm E1E1}_{nd,1s}(\omega)=e^4\frac{16\,\omega^3\omega'^3}{135\pi}
\left|S_{1s,nd}(\omega)+S_{1s,nd}(\omega')\right|^2d\omega\,,
\end{eqnarray}
with notations
\begin{eqnarray}
\label{24}
S_{1s,ns/nd}(\omega)=\sum\limits_{n'p}
\frac{\langle R_{1s}|r|R_{n'p}\rangle\langle R_{n'p}|r|R_{ns/nd}\rangle}{E_{n'p}-E_{ns}+\omega}\,,
\end{eqnarray}
\begin{eqnarray}
\label{25}
\langle R_{n'l'}|r|R_{nl}\rangle=\int\limits_{0}^{\infty}r^3R_{n'l'}(r)R_{nl}(r)dr\,,
\end{eqnarray}
where $R_{nl}(r)$ are the radial parts of the nonrelativistic hydrogenic wave
functions, and $E_{nl}$ are the corresponding energies.

The corresponding decay rate for the two-photon transitions
$ns/nd \rightarrow 1s$ can be obtained by integration of Eq. (\ref{21})
[in nonrelativistic case Eqs. (\ref{22}) and (\ref{23})] over the entire frequency interval
\begin{eqnarray}
\label{26}
W^{\rm E1E1}_{AA'}=\frac{1}{2}
\int\limits_0^{\omega_0}dW^{\rm E1E1}_{AA'}(\omega)\,,
\end{eqnarray}
where $\omega_0 = E_A - E_A'$.
The gauge invariance of the transition probability was investigated in \cite{Grant}
for the one-photon transitions. The two different forms for the ${\rm E}k$ one-photon
probabilities in combination with different gauges were obtained in \cite{Nikitin}.
In our paper \cite{labzowsky2006} the results of \cite{Nikitin} were
applied for the evaluation of the two-photon transition probabilities.
The gauge invariance serves as a tool for testing the numerical evaluation procedure
in both relativistic and nonrelativistic case.

The aim of this paper is the investigation of the nonresonant two-photon decay
rates from the highly excited hydrogenic states, since these transitions
could play an important role in studies of the anisotropy of the
cosmic microwave background. It was expected that two-photon transitions
from the highly excited states could contribute essentially in the
recombination process. Accordingly, we study the behaviour of the frequency
distribution function of the transition probability of the initial state with
the increasing principal quantum number $n$. Comparison between the relativistic and
nonrelativistic calculations are performed in order to examine possible
deviation from the dipole approximation.
The results of numerical calculations are presented in Table~\ref{tab:1} for
the $ns - 1s$ two-photon E1E1 transitions.
For the analysis of the $n$-behaviour of the ``pure'' two-photon decays we
present the results in terms the normalized distribution function
\begin{eqnarray}
\label{27}
Q^{\rm E1E1}_{ns,1s}=\frac{1}{2}\frac{dW^{\rm E1E1}_{ns,1s}}{W^{\rm E1E1}_{2s,1s}d\omega}\,n^3
\end{eqnarray}
at the equal frequencies of the emitted photons $x = \omega/\omega_0 = 0.5$.
The choice of the frequency is caused by the absence of the cascades in this region.
The E1E1 two-photon transition probability $W^{\rm E1E1}_{2s,1s}=8.229$ s$^{-1}$
is taken as normalization factor. Moreover, the normalized distribution function
is multiplied by factor $n^3$ in order to emphasize the dominant behaviour of
$Q^{\rm E1E1}_{ns,1s}$ for large $n$ values.
In Table~\ref{tab:1} we present also the corresponding quantity $Q^{\rm E1E1}_{nd,1s}$
for the E1E1 two-photon $nd - 1s$ transitions. Again the normalization factor
$W^{\rm E1E1}_{2s,1s}$ for the frequency distribution function is taken.

These results reveal that for large values of $n$ the scaling of the frequency
distribution function for the E1E1 two-photon $ns/nd - 1s$ transitions is close
to $1/n^3$ and the dipole approximation holds even for the large $n$.
This means the decrease of the nonresonant two-photon transitions contribution for
large values of $n$ to the recombination dynamics of the primordial hydrogen
in the Universe. The relativistic corrections appear to be not essential for
the considering two-photon transitions from the highly excited levels to the
ground state.
The Table~\ref{tab:1} shows also, that two-photon E1E1 $ns - 1s$ transition probabilities
decrease faster then $1/n^3$ for the low values $n$ and picture changes for the
$n\approx 10$ when the normalized
quantity $Q^{\rm E1E1}_{ns,1s}$ reaches the asymptotic value $\approx 6.6$. For the
$nd - 1s$ E1E1 two-photon transitions behaviour is quite different. Namely, for the
$n\leqslant 40$ the values of the transition rates decrease slower $1/n^3$ and then,
as in the case of $ns - 1s$ transitions, reach the $1/n^3$ asymptotics.
\begin{table}
\caption
{Comparison between the results of relativistic and nonrelativistic
calculations of the normalized distribution functions $Q^{\rm E1E1}_{ns,1s}$
and $Q^{\rm E1E1}_{nd,1s}$ at the equal frequencies of the emitted photons
$x = 0.5$ for the interval of the principal quantum number $n=[2,100]$ and
$n=[3,100]$, respectively.}
\label{tab:1}
\begin{tabular}{|| l | c | c | c | c ||}
\hline \hline
$n$ & $Q^{\rm E1E1}_{ns,1s}({\rm nonrel.})$ & $Q^{\rm E1E1}_{ns,1s}({\rm rel.})$
        & $Q^{\rm E1E1}_{nd,1s}({\rm nonrel.})$ & $Q^{\rm E1E1}_{nd,1s}({\rm rel.})$ \\ \hline
  2 & 10.35  & 10.35  & ---   & ---    \\
  3 &  8.527 &  8.528 & 32.21 & 32.21  \\
  5 &  7.255 &  7.255 & 49.97 & 49.97  \\
  8 &  6.674 &  6.765 & 56.39 & 56.46  \\
 10 &  6.591 &  6.648 & 57.97 & 58.00  \\
 20 &  6.595 &  6.490 & 60.13 & 60.08  \\
 30 &  6.576 &  6.460 & 60.51 & 60.46  \\
 40 &  6.567 &  6.450 & 60.65 & 60.60  \\
 50 &  6.562 &  6.445 & 60.71 & 60.66  \\
 60 &  6.559 &  6.442 & 60.71 & 60.70  \\
 70 &  6.558 &  6.441 & 60.73 & 60.72  \\
 80 &  6.557 &  6.439 & 60.74 & 60.73  \\
 90 &  6.556 &  6.439 & 60.74 & 60.74  \\
100 &  6.556 &  6.438 & 60.75 & 60.75  \\
\hline \hline
\end{tabular}
\end{table}

To make the picture complete the relativistic evaluations have been also
performed for E2E2, M1M1 and E1M2 transitions as the corrections to the E1E1
$ns$-level decay. Their contribution are compiled in Table~\ref{tab:2}
and appear to be negligible.
\begin{table}
\caption{
The relativistic values of the normalized distribution functions
$Q^{\rm E2E2}_{ns,1s}$, $Q^{\rm M1M1}_{ns,1s}$ and $Q^{\rm E1M2}_{ns,1s}$
at the equal frequencies of the emitted photons $x = 0.5$ for the interval
of the principal quantum number $n=[2,100]$.}
\label{tab:2}
\begin{tabular}{|| l | c | c | c ||}
\hline \hline
$n$ & $Q_{ns,1s}^{\rm E2E2}({\rm rel.})$ & $Q_{ns,1s}^{\rm M1M1}({\rm rel.})$ 
         & $Q_{ns,1s}^{\rm E1M2}({\rm rel.})$\\ \hline
  2 & $1.168 \times 10^{-11}$ & $3.019 \times 10^{-11}$ & $ 2.581 \times 10^{-10}$ \\
  5 & $1.907 \times 10^{-11}$ & $5.277 \times 10^{-11}$ & $ 2.962 \times 10^{-10}$ \\
 10 & $2.967 \times 10^{-11}$ & $5.620 \times 10^{-11}$ & $ 2.886 \times 10^{-10}$ \\
 20 & $3.265 \times 10^{-11}$ & $5.916 \times 10^{-11}$ & $ 2.861 \times 10^{-10}$ \\
 40 & $3.342 \times 10^{-11}$ & $5.932 \times 10^{-11}$ & $ 2.853 \times 10^{-10}$ \\
 60 & $3.356 \times 10^{-11}$ & $5.932 \times 10^{-11}$ & $ 2.853 \times 10^{-10}$ \\
 80 & $3.361 \times 10^{-11}$ & $5.932 \times 10^{-11}$ & $ 2.852 \times 10^{-10}$ \\
 90 & $3.363 \times 10^{-11}$ & $5.932 \times 10^{-11}$ & $ 2.851 \times 10^{-10}$ \\
100 & $3.364 \times 10^{-11}$ & $5.932 \times 10^{-11}$ & $ 2.851 \times 10^{-10}$ \\
\hline \hline
\end{tabular}
\end{table}
The Table~\ref{tab:2} shows that the contributions of the higher multipoles
to the nonresonant two-photon emission for the $ns - 1s$ decays is negligible.
The parametric estimation for these transition rates is $(\alpha Z)^{10}$ in atomic
units. The values of the M1M1, E2E2, and E1M2 transition
probabilities are in good accordance with the reported in \cite{amaro2009}.
It is obvious that corresponding values of the transition probabilities for the
two-photon emission of the $nd$ states are also negligible and all of them can be
omitted in astrophysical investigations.

\section{E1E2 decay of the $np$ states}
In this section we evaluate the two-photon E1E2 transition rates
$np \rightarrow 1s$. Relativistic and nonrelativistic calculations for
the $2p-1s$ E1E2 transition were previously performed in papers
\cite{labzowsky2005} and \cite{labzowsky2006}. We will follow the paper
\cite{labzowsky2006}, where two-photon E1E2 transition probability was
evaluated in different forms and gauges. For reasons of simplicity in this
section we present the evaluation of the transition rate within the nonrelativistic
length form. The electric multipole operators can be written as (nonrelativistic limit):
\begin{eqnarray}
V^{{\rm E}k}(\omega)=\sqrt{\frac{k+1}{k}}
\frac{2\omega^{k+\frac{1}{2}}}{(2k+1)!!}r^kY_{k\,-M}\,,
\end{eqnarray}
where $Y_{k\,-M}$ is a spherical harmonic.
Accordingly, the two-photon decay rate of the atomic
state $A$  with the emission of two electric photons can be written as
(see, e.g., \cite{labzowsky2006})
\begin{eqnarray}
dW_{A A'}^{{\rm E}k{\rm E}k'}(\omega,\omega')=\sum\limits_{M M' m_A m_{A'}}\left|
  \sum\limits_N\frac{(A'|V^{{\rm E}k}(\omega)|N)(N|V^{{\rm E}k'}(\omega')|A)}{E_N-E_A+\omega'}
+ \sum\limits_N\frac{(A'|V^{{\rm E}k'}(\omega')|N)(N|V^{{\rm E}k}(\omega)|A)}{E_N-E_A+\omega}
\right.
\nonumber\\
\left.
+ \sum\limits_N\frac{(A'|V^{{\rm E}k}(\omega')|N)(N|V^{{\rm E}k'}(\omega)|A)}{E_N-E_A+\omega}
+ \sum\limits_N\frac{(A'|V^{{\rm E}k'}(\omega)|N)(N|V^{{\rm E}k}(\omega')|A)}{E_N-E_A+\omega'}
\right|^2\delta
\left(\omega+\omega'-E_A+E_{A'}\right)\, d\omega d\omega'\,.
\end{eqnarray}
To perform the summation over the complete set of intermediate states we employ the explicit
expression for the Coulomb Green Function \cite{labzowsky2003}. With the aid of eigenmode decomposition
of the Coulomb Green function the probability of the two-photon decay process in
nonrelativistic limit takes the form \cite{labzowsky2006}
\begin{eqnarray}
dW_{2p,1s}^{\rm E1E2}(\omega) =
\alpha^6\frac{2^2\omega^3\omega'^3}{3^35^2\pi}\left[\omega'^2
\left|I_1(\omega')+I_2(\omega)\right|^2+\omega^2
\left|I_1(\omega)+I_2(\omega')\right|^2\right]d\omega \,,
\end{eqnarray}
where
\begin{eqnarray}
I_1(\omega)=\frac{1}{\sqrt{6}}\int\limits_0^{\infty}
\int\limits_0^{\infty}dr_1dr_2\,r_1^3r_2^5\,e^{-r_1-\frac{r_2}{2}}\,
g_1(E_A-\omega;r_1, r_2)
\end{eqnarray}
and
\begin{eqnarray}
I_2(\omega)=\frac{1}{\sqrt{6}}\int\limits_0^{\infty}
\int\limits_0^{\infty}dr_1dr_2\,r_1^4r_2^4\,e^{-r_1-\frac{r_2}{2}}\,
g_2(E_A-\omega;r_1, r_2)\,.
\end{eqnarray}
The decomposition of radial part of the Coulomb Green function reads
\begin{eqnarray}
\label{green}
g_l(\nu; r,r')=\frac{4Z}{\nu}\left(\frac{4}{\nu^2}rr'\right)^l
\exp\left(-\frac{r+r'}{\nu}\right)\sum\limits_{n=0}^{\infty}
\frac{n!L^{2l+1}_n\left(\frac{2r}{\nu}\right)L^{2l+1}_n
\left(\frac{2r'}{\nu}\right)}{(2l+1+n)!(n+l+1-\nu)}\,,
\end{eqnarray}
where $L^{2l+1}_n$ are the generalized Laguerre polynomials.
The corresponding radial integrals can be evaluated analytically.
Equations (22) and (23) can be easily generalized for the case of
$np\rightarrow \gamma(E1)+\gamma(E2)+1s$ transitions by the replacing the
radial function $2p$ state, which is equal to
$R_{21}(r)=\frac{r}{2\sqrt{6}}e^{-r/2}$, by the required one.

Finally, integrating over frequencies $\omega$ yields ($\omega_0=E_{2p}-E_{1s}$)
\begin{eqnarray}
\label{4.9}
W^{\rm E1E2}_{2p,1s}=\frac{1}{2}\int\limits_0^{\omega_0} dW_{2p,1s}^{\rm E1E2}(\omega)
= 1.98896 \times 10^{-5}\,(\alpha Z)^8\, {\rm a.u.} =
6.61197 \times 10^{-6}\, {\rm s}^{-1}\,(Z=1)\,.
\end{eqnarray}
The $Z$-dependence of the $W^{\rm E1E2}_{2p,1s}$
transition probability is also indicated. Compared with the relativistic result
the relative discrepancy is about $0.1\%$. The parametric estimation of the transition
rate $W^{\rm E1E2}_{2p,1s}$ is $(\alpha Z)^8$ in atomic units. This is $\alpha^2$
times less than the corresponding parametric estimate for the E1E1 transition
probability, which is $(\alpha Z)^6$. However, with respect to the achieved
accuracy of the astrophysical experiments, the E1E2 transition rates could be
considered as a correction to the two-photon processes.

In Table~\ref{tab:3} the relativistic results for the
$np_{1/2}\rightarrow \gamma(E1)+\gamma(E2)+ (n-1)s$ transition rates are presented.
Good agreement with the corresponding nonrelativistic results is found.
Bad convergence properties of the finite basis set representation restrict
our calculations of the transitions between neighbouring states to $n = 5$.
This can be checked by considering the level of accuracy up to which gauge
invariance is preserved in the numerical evaluations.
For the $nl \rightarrow 2\gamma + 1s$ transitions the gauge invariance is preserved
with good accuracy up to the $n=100$. In Table~\ref{tab:3} we also display the
relativistic and nonrelativistic values
for the E1E2 normalized distribution function
\begin{eqnarray}
Q^{\rm E1E2}_{np_{1/2},\,1s} = \frac{1}{2}\frac{dW^{\rm E1E2}_{np_{1/2},\,1s}}
{W^{\rm E1E2}_{2p_{1/2},\,1s}d\omega}\,n^3
\end{eqnarray}
at the equal frequencies of the emitted photons $x = \omega/\omega_0 = 0.5$.
One can see from the table that the behaviour of
the nonresonant E1E2 two-photon emission with the increasing
$n$ values is the same as for E1E1 probability of the $ns/nd$-states emission.
And again the nonrelativistic dipole approximation works well.

\section{E1M1 decay of the $np$ states}
A parametric estimate shows that the E1M1 transition probability is of the same
order of magnitude as for E1E2, therefore, it should be included in our
consideration. For the first time the E1M1 transition rate for the
$2p_{1/2} \rightarrow 1s$
emission process was evaluated in \cite{labzowsky2005,labzowsky2006}.
Our results were later confirmed by Amaro et al. in \cite{amaro2009}.

The expression for the E1M1 two-photon transition probability is given by
\begin{eqnarray}
\label{E1M1i}
dW_{A A'}^{\rm E1M1}(\omega)=\sum\limits_{M_{\rm E} M_{\rm M} m_A m_{A'}}
\left|\sum\limits_{N}\frac{(A'|V^{\rm
E1}(\omega)|N)(N|V^{{\rm M1}}(\omega')|A)}{E_N-E_A+\omega'}
+\sum\limits_{N}\frac{(A'|V^{\rm M1}(\omega')|N)(N|V^{\rm E1}(\omega)|A)}{E_N-E_A+\omega}
 \right.\nonumber\\
 \left.
+\sum\limits_{N}\frac{(A'|V^{\rm E1}(\omega')|N)(N|V^{\rm M1}(\omega)|A)}{E_N-E_A+\omega}
+\sum\limits_{N}\frac{(A'|V^{\rm M1}(\omega)|N)(N|V^{\rm E1}(\omega')|A)}{E_N-E_A+\omega'}
\right|^2\,d\omega\,.
\end{eqnarray}
In the nonrelativistic limit the corresponding magnetic dipole operator reads
$V^{\rm M1}(\omega)=\sqrt{\frac{4}{3}}\mu_0\omega^{3/2}
\left(\hat{j}_{1M_{\rm M}}+\hat{s}_{1M_{\rm M}}\right)$, where $\mu_0=\alpha/2$
is the Bohr magneton, $\hat{j}_{1M_{\rm M}}$ and $\hat{s}_{1M_{\rm M}}$ are the
spherical components of the total angular momentum and the spin operator
(spherical tensors of rank 1) of the electron. Since the operator for the magnetic
photon includes total angular momentum and spin operator, coupled wave functions
characterized by the set of quantum numbers $N=\{nlsjm\}$ should be used, i.e.
\begin{eqnarray}
\phi_{nlsjm}=\sum\limits_{m_lm_s}C^{jm}_{lm_l\,sm_s}R_{nl}(r)Y_{m_l}^{(l)}({\bf n}_{\bf r})
\chi_{sm_s}\,,
\end{eqnarray}
where $\chi_{sm_s}$ $(s=1/2)$ is the spin function. The magnetic potentials in
Eq.~(\ref{E1M1i}) do not depend on radial variables. Thus, only the intermediate
states with $nl=n_Al_A$ or $nl=n_{A'}l_{A'}$ will contribute to the transition probability
in Eq.~(\ref{E1M1i}). Performing angular integrations and summations over all
projections one arrives  at the expressions
\begin{eqnarray}
dW_{2p_{1/2},\,1s}^{\rm E1M1}(\omega)=\alpha^6\frac{2^8\mu_0^2}{\pi}
\left(\frac{2}{3}\right)^{12}\omega\omega'^3d\omega
\end{eqnarray}
and
\begin{eqnarray}
W_{2p_{1/2},\,1s}^{\rm E1M1}=\frac{1}{2}\int\limits_0^{\omega_0}
dW_{2p_{1/2},\,1s}^{\rm E1M1}(\omega)=
\alpha^8\frac{2^5}{\pi}\left(\frac{2}{3}\right)^{12}
\int\limits_0^{\omega_0}\omega\left(\omega_0-\omega\right)^{3}d\omega\,,
\end{eqnarray}
with $\omega_0 = E_{2p_{1/2}}-E_{1s}$.
As the final result we obtain (see also \cite{labzowsky2006})
\begin{eqnarray}
W_{2p_{1/2},\,1s}^{\rm E1M1}=\frac{(\alpha Z)^8}{10935\pi}\, {\rm a.u.} =
9.6769\times 10^{-6}\, {\rm s}^{-1}\,(Z=1)\,.
\end{eqnarray}
Again the $Z$-dependence of the $W^{\rm E1M1}_{2p_{1/2},1s}$ transition
probability is explicitly indicated. Comparison with the result of a fully relativistic
calculation now reveals a deviation of about $0.1\%$.

In case of arbitrary principal quantum number $n$ due to the condition
\begin{eqnarray}
\langle R_{nl}|M1|R_{n'l'}\rangle\sim \delta_{nn'}\delta_{ll'}
\end{eqnarray}
only the diagonal matrix elements remain in the sum over the intermediate states
in Eq. (\ref{E1M1i}). Therefore, in the nonrelativistic limit no cascades will arise 
in the considering processes involving the magnetic dipole photons.
We should note that, in the relativistic case the cascades will arise,
however, the contribution of the cascades appears to be negligible
in the case of hydrogen atom.

For the E1M1 transition probability we have
\begin{eqnarray}
\label{e1m1-1}
dW_{np_{1/2},\,ls}^{\rm E1M1}(\omega) = \frac{\alpha^8}{2\pi}\left(\frac{2}{3}\right)^3
\omega\omega'(\omega^2+\omega'^2)
\left[\int\limits_0^{\infty}R_{np}(r)r^3R_{ls}(r)dr\right]^2d\omega\,.
\end{eqnarray}
After the integration over radial variable $r$ and the frequency $\omega$
within the interval $[0,E_{np_{1/2}}-E_{1s}]$ we obtain
\begin{eqnarray}
\label{e1m1-2}
W_{np_{1/2},\,1s}^{\rm E1M1} = \frac{8}{135\pi n^3}
\left(\frac{n-1}{n+1}\right)^{2n}(\alpha Z)^8\,.
\end{eqnarray}
For $n\gg 1$ it follows
$\displaystyle W^{\rm E1M1}_{np_{1/2},\,1s} \approx \frac{8}{135\pi}\left(1-\frac{4}{3n^2}\right)\frac{(\alpha Z)^8}{\exp^4n^3}$,
i.e. the same scaling law arises as for E1E1 transitions.

For the $np_{1/2}\rightarrow \gamma(E1)+\gamma(M1)+2s$ emission process
we obtain in the same way
\begin{eqnarray}
\label{e1m1-3}
W_{np_{1/2},\,2s}^{\rm E1M1} = 
\frac{4}{135\pi n^3}\left(\frac{n-2}{n+2}\right)^{2n}\frac{(n^2-1)}{(n^2-4)}(\alpha Z)^8\,.
\end{eqnarray}
Similarly, for $n\gg 1$ the scaling is
$\displaystyle W^{\rm E1M1}_{np_{1/2},\,2s} \approx \frac{4}{135\pi}\left(1-\frac{32}{3n^2}\right)\frac{(\alpha Z)^8}{\exp^8n^3}$.

In Table~\ref{tab:3} the relativistic results for the
$np_{1/2} \rightarrow \gamma(E1)+\gamma(M1)+ (n-1)s$ transition rates are presented.
These results are in a good agreement with the corresponding nonrelativistic values,
which can be easily obtained in terms of Eqs.~(\ref{e1m1-1})-(\ref{e1m1-3}).
Again we restrict ourselves to $n = 5$ for the transitions between neighbouring states
[$np_{1/2}$ and $(n-1)s$].
In Table~\ref{tab:3} we also display the relativistic and nonrelativistic
values for the E1M1 normalized distribution function
\begin{eqnarray}
Q^{\rm E1M1}_{np_{1/2},\,1s} = \frac{1}{2}\frac{dW^{\rm E1M1}_{np_{1/2},\,1s}}
{W^{\rm E1M1}_{2p_{1/2},\,1s}d\omega}\,n^3
\end{eqnarray}
at the equal frequencies of the emitted photons $x = \omega/\omega_0 = 0.5$.
As in previous cases the behaviour of the nonresonant E1M1 two-photon emission
is $1/n^3$. The nonrelativistic dipole approximation holds in this case as well.
\begin{table}
\caption{The relativistic values of the transition probabilities
$W^{\rm E1E2}_{np_{1/2},\,(n-1)s}$ and $W^{\rm E1M1}_{np_{1/2},\,(n-1)s}$ in units s$^{-1}$
are presented in second and fifth columns, respectively. In the third and fourth
columns the relativistic and nonrelativistic values for the distribution function
$Q^{\rm E1E2}_{np_{1/2},\,1s}$ are given at the equal frequencies of the emitted
photons $x = 0.5$. In the sixth and seventh columns the relativistic and nonrelativistic
values for the distribution function $Q^{\rm E1M1}_{np_{1/2},\,1s}$ are presented
at $x = 0.5$.
}
\label{tab:3}
\begin{tabular}{| l | c | c | c || c | c | c |}
\hline \hline
$n$ & $W^{\rm E1E2}_{np_{1/2},\,(n-1)s}$ & $Q^{\rm E1E2}_{np_{1/2},1s}({\rm rel.})$ &
 $Q^{\rm E1E2}_{np_{1/2},1s}({\rm nonrel.})$ & $W^{\rm E1M1}_{np_{1/2},\,(n-1)s}$ &
  $Q^{\rm E1M1}_{np_{1/2},1s}({\rm rel.})$ & $Q^{\rm E1M1}_{np_{1/2},1s}({\rm nonrel.})$ \\ \hline
  2 & $6.612 \times 10^{-06}$ & 15.31 & 15.32 & $9.682 \times 10^{-06}$ &  9.99 & 13.33 \\
  3 & $3.737 \times 10^{-08}$ & 3.31  &  3.32 & $1.189 \times 10^{-08}$ & 12.65 & 14.24 \\
  4 & $9.753 \times 10^{-10}$ & 14.26 & 14.27 & $1.989 \times 10^{-10}$ & 13.60 & 14.51 \\
  5 & $6.221 \times 10^{-11}$ & 21.91 & 21.91 & $1.025 \times 10^{-11}$ & 14.04 & 14.63 \\
 10 &        ------           & 34.64 & 34.53 &        ------           & 14.63 & 14.78 \\
 20 &        ------           & 38.28 & 38.45 &        ------           & 14.78 & 14.82 \\
 30 &        ------           & 38.94 & 39.13 &        ------           & 14.81 & 14.83 \\
 40 &        ------           & 39.36 & 39.22 &        ------           & 14.82 & 14.83 \\
 50 &        ------           & 39.33 & 39.47 &        ------           & 14.82 & 14.83 \\
 60 &        ------           & 39.39 & 39.52 &        ------           & 14.82 & 14.83 \\
 70 &        ------           & 39.43 & 39.56 &        ------           & 14.82 & 14.83 \\
 80 &        ------           & 39.46 & 39.58 &        ------           & 14.82 & 14.83 \\
 90 &        ------           & 39.47 & 39.60 &        ------           & 14.83 & 14.83 \\
100 &        ------           & 39.48 & 39.61 &        ------           & 14.83 & 14.84 \\
\hline \hline
\end{tabular}
\end{table}

\section{Conclusions}
Relativistic and nonrelativistic calculations have been performed and compared.
The numerical evaluation have been carried out employing the dual-kinetic-balance
finite basis set method \cite{shabaev:2004:130405} with basis functions constructed
from B-splines \cite{sapirstein:1996:5213}. For the nonrelativistic calculations the
analytical expression for the nonrelativistic Coulomb Green function has been utilized.

We have proved that the dependence of the transition probabilities on the principal
quantum number of the initial state $n$ out of the cascade regions is close to $1/n^3$.
Therefore, the contribution of the highly excited states turn out to be much less
significant for the astrophysical purposes as expected.
Previously, the $n$-dependence of the nonresonant contribution was estimated
in \cite{dubrovich2005,chluba2008}. It was found that the nonresonant transition
rates scale
roughly linear, when increasing towards larger $n$.
The nonresonant two-photon rates was defined, e.g., in \cite{chluba2008},
via neglecting the resonant state in the summation over the entire spectrum of
the intermediate states.
However, the term with the resonant state alone contributes both to the resonant
rate and to the nonresonant one. Therefore, to our mind,
the investigation of the $n$-behaviour of the nonresonant emission
should be performed by analyzing the differential transition rate
in the region, where the contribution of the cascades is negligible.

The relativistic calculations were performed aiming for the search of the influence
of effects
beyond the nonrelativistic dipole approximation. The gauge invariance served as
a check of the calculations. The nonrelativistic calculations were performed in
the length gauge. In principle, the gauge invariance also could be used in this case
(see, for example, \cite{labzowsky2006}), but comparison with the corresponding
relativistic values is enough for our purposes here.
We have compared the relativistic and nonrelativistic results to understand whether
it is necessary to go beyond the dipole approximation. It was expected, that for the
highly excited states ($n\gg 1$) the dipole approximation might be not very accurate
due to the large argument of the Bessel function. However, we were able to show
that even for the $n = 100$ the deviation between the nonrelativistic dipole 
approximation and relativistic theory is not significant. The relative difference
for the corresponding values does not exceed $1.8\%$.
Since we always consider the two-photon decays into the ground state,
this can be the reason that the dipole approximation is valid even
for such high values of $n$. The presence of the ground state provides a short-range
cutoff for one of the radial variable ($r$, for instance, in Eq.~(\ref{green})),
then the range of second radial variable $r'$ will also be small,
because of the exponential suppression factor $\exp[-(r+r')]$ coming from the Green
function (\ref{green}).
Moreover, we have evaluated M1M1, E2E2, and E1M2 $ns-1s$ transition rates.
It was shown that they behave like $1/n^3$ also and the parametric estimation for
them is $(\alpha Z)^{10}$ in atomic units.

On the score of the highly excited states are fully mixed
we have considered also $np-1s$ two-photon transitions.
In the frame of the present work we have evaluated the
E1E2 and E1M1 two-photon transitions rates for the $np - 1s$ emission processes
and investigated the behaviour of the nonresonant decay rates as function of the
principal quantum number $n$ of the initial state.
For large values of $n$ the nonresonant emission of the two-photon E1E2 and E1M1
emission is proportional to $1/n^3$.
The nonrelativistic dipole approximation is again in a good agreement
with the relativistic calculations.
Within the framework of the relativistic approach
we have evaluated  also the total
decay rates for the $np \rightarrow (n-1)s$ E1E2 and E1M1 transitions up to $n = 5$,
and compared them with the nonrelativistic results.
The restriction to $n = 5$ is set by the level of accuracy upto which the gauge invariance
is achieved in the numerical evaluations.
The relativistic calculations based on the finite basis set method does
not converge good in this case and others methods should be employed.
Moreover, we have derived analytically simple formulas for the two-photon
$np\rightarrow \gamma(E1)+\gamma(M1)+1s$ and $np\rightarrow \gamma(E1)+\gamma(M1)+2s$
emission processes within the nonrelativistic approach. It has been shown that
in the presence of one magnetic dipole photon in the two-photon emission process
the transition probability does not contain cascade contribution in the
nonrelativistic limit. Account for the relativistic corrections leads to appearance
of the cascade contributions. However, the smallness of the two-photon
E1E2 and E1M1 transitions leaves them behind the astrophysical investigations.

The main conclusion that we can stay is the rapid reduction for the all considered
nonresonant two-photon transition rates with the increase of $n$.
The dependence $1/n^3$ has been established for ``pure'' two-photon transition rates,
corresponding to the radiation escape from the interaction with the matter.
It was shown that the nonrelativistic dipole consideration of the two-photon transition
rates from the highly excited states is valid even for large $n$, in case that the
final state is one of the lowest state of the atom.

\begin{center}
Acknowledgments
\end{center}
We thank the referees for helpful remarks.
The authors acknowledge financial support from DFG and GSI. The work was also
supported by RFBR (grant Nr. 08-02-00026). The work of D.S. was supported by
the Non-profit Foundation ``Dynasty'' (Moscow). D.S. and L.L. acknowledge
the support by the Program of development of scientific potential of High School,
Ministry of Education and Science of Russian Federation (grant Nr. 2.1.1/1136).
The work of A.V.V. was supported by the DFG (grant Nr. VO 1707/1-1).

\end{document}